\documentclass[prb,twocolumn,floats]{revtex4}
\usepackage{amsmath}
\usepackage{amsfonts}
\usepackage{graphicx}
\newcommand{\rr}{{\bf r}}
\newcommand{\kk}{{\bf k}}
\newcommand{\KK}{{\bf K}}
\newcommand{\RR}{{\bf R}}
\newcommand{\qq}{{\bf q}}
\begin{document}
\title{Role of CuO chains in vortex core structure in
$\mathrm{YBa_2Cu_3O_{7-\delta}}$} 
\author{W. A. Atkinson$^1$ and J. E. Sonier$^{2,3}$}
\affiliation{{}$^1$ Department of Physics and Astronomy, Trent
University, 1600 West Bank Dr., Peterborough ON, K9J 7B8, Canada \\
{}$^2$Department of Physics, Simon Fraser University, 8888 University
Dr., Burnaby BC, V5A 1S6, Canada \\
{}$^3$ Canadian Institute for Advanced Research, 180 Dundas Street West,
Toronto ON M5G 1Z8, Canada}
\date{\today}

\begin{abstract}
The Bogoliubov-deGennes equations are solved for a proximity model for
$\mathrm{YBa_2Cu_3O_{7-\delta}}$ in a magnetic field.
The model explicitly includes the effects
of the one-dimensional CuO chains, whose influence on the vortex core
structure is studied.
The rapid vortex core contraction as a function of field which is seen
experimentally at low magnetic fields is naturally explained by the
presence of the chains.
\end{abstract}
\maketitle

\section{Introduction}

As first observed by Golubov and Hartmann \cite{Golubov1994} using
scanning tunneling microscopy (STM), the vortex cores in NbSe$_2$
shrink with increasing magnetic field.  Subsequent muon spin rotation
($\mu$SR) experiments confirmed this to be the case in NbSe$_2$
\cite{Sonier1997a} but also in CeRu$_2$ \cite{Kadono2001}, YNi$_2$B$_2$C
\cite{Ohishi2002}, LuNi$_2$B$_2$C \cite{Price2002}, V$_3$Si
\cite{Sonier2004a}, V \cite{Laulajainen2006}, Nb$_3$Sn \cite{Kadono2006}
and YBa$_2$Cu$_3$O$_{7-\delta}$
\cite{Sonier1997b,Sonier1999a,Sonier1999b,Sonier2007}.  The STM
experiments probe the spatial variation of the local density of
states, whereas $\mu$SR is sensitive to the spatial dependence of the
local internal magnetic field $B(r)$. The vortex core size is
determined from the $\mu$SR measurements by fitting to a theoretical
function for $B(r)$ that includes a cutoff function $F({\bf G}, \xi)$,
where {\bf G} are the reciprocal lattice vectors and $\xi$ is the
superconducting coherence length.  The functional form of $F({\bf G},
\xi)$ depends on the spatial dependence of the superconducting order
parameter $\Delta(r)$ in the core region. Since there is no way of
knowing exactly what this is in a real material, the fitted value of
$\xi$ reflects differences between the theoretical model and the real
spatial dependence of the local field about the vortex cores.
Consequently, $\xi$ is generally not the coherence length, but rather
a measure of the vortex core size. A second definition of the vortex
core size is the radius $r_0$ at which the supercurrent density
$|j(r)|$ calculated from $B(r)$ reaches a maximum. While this
definition is robust with respect to the assumed model for $B(r)$,
there is a contribution to the field dependence of $r_0$ that comes
naturally from the overlap of the $j(r)$ profiles of neighboring
vortices \cite{Sonier2004b}.

Kogan and Zhelezina \cite{Kogan2005} have proposed a model based on
weak-coupling BCS theory that explains the field dependence of the
core size in clean high-$\kappa$ superconductors as being due to a
field-dependent superconducting coherence length. Their model
qualitatively describes the $\mu$SR results for CeRu$_2$, NbSe$_2$,
V$_3$Si and YNi$_2$B$_2$C.  A field-dependent coherence length has
also been suggested to be the source of the anomalous
field-independent flux-line lattice form factor observed in
small-angle neutron scattering measurements on CeCoIn$_5$
\cite{DeBeer2006}.  However, Ichioka and Machida \cite{Ichioka2007}
have recently argued that this is caused by paramagnetic moments due
to Zeeman splitting of the Fermi surfaces for spin-up and spin-down
electrons.

Within the framework of the microscopic theory, the field dependence
of the vortex core size can be explained without invoking a
field-dependent coherence length. Solutions of the quasiclassical
Usadel equations for a dirty $s$-wave superconductor
\cite{Golubov1994,Sonier1997a}, and solutions of the quasiclassical
Eilenberger equations for clean $s$-wave and $d$-wave superconductors
\cite{Ichioka1999a,Ichioka1999b} show that the field dependences of
the electronic and magnetic structures of the vortex cores are
coupled. As explained in
Refs.~[\onlinecite{Ichioka1999a,Ichioka1999b}], the shrinking of the
vortex cores with increasing $H$ occurs due to an increased overlap of
the wave functions of the quasiparticle core states from
nearest-neighbor vortices.  This delocalization of quasiparticles,
beginning with the more spatially extended wave functions of the
higher-energy core states, increases the slope of $\Delta(r)$ near $r
= 0$, which corresponds to a reduction in the size of the vortex core.
Experimentally, this picture is strongly supported by the remarkable
correlation found in V$_3$Si \cite{Sonier2004a} and NbSe$_2$
\cite{Callaghan2005} between the field dependences of the core size
and the electronic thermal conductivity.

As pointed out in Ref.~[\onlinecite{Sonier2004b}] the $\mu$SR
measurements of NbSe$_2$ and YBa$_2$Cu$_3$O$_{7-\delta}$ are unusual
in that at low fields the core size $\xi$ exceeds the value of the
coherence length calculated from the upper critical field
$H_{c2}$. Vortex cores larger than estimated from $H_{c2}$ have also
been observed at low field by STM on the $\pi$-band of the two-gap
superconductor MgB$_2$ \cite{Eskildsen2002}.  While superconductivity
on both the $\pi$ and $\sigma$ bands of MgB$_2$ contribute to the
electronic structure of the vortex cores, at low field the dominant
contribution comes from the loosely bound quasiparticle core states
associated with the smaller gapped $\pi$-band \cite{Nakai2002}.  With
increasing $H$ these core states rapidly delocalize so that at high
field the core size, and hence $H_{c2}$, is determined by the
intrinsic superconductivity on the $\sigma$-band. Like MgB$_2$, there
is experimental evidence for distinct energy gaps on different Fermi
sheets in NbSe$_2$ \cite{Yokoya2001,Boaknin2003,Rodrigo2004}.
Recently, the effects of the Fermi-surface sheet dependent
superconductivity on the vortex core size became discernible in a
low-temperature $\mu$SR study of NbSe$_2$ \cite{Callaghan2005}. In the
same spirit, one of us suggested that the large vortex cores at low
field in YBa$_2$Cu$_3$O$_{7-\delta}$ may be caused by the occurrence
of superconductivity on the CuO chain bands.\cite{Sonier2004b}

While all of the cuprate high temperature superconductors are based
around conducting two-dimensional CuO$_2$ layers, 
$\mathrm{YBa_2Cu_3O_{7-\delta}}$ and
$\mathrm{YBa_2Cu_4O_{8}}$ are unique among the cuprates in having an
additional type of conducting layer, made of one-dimensional CuO
chains.  Band structure calculations\cite{Andersen1995} and recent
photoemission
experiments\cite{Schabel1998,Lu2001,Zabolotnyy2006,Nakayama2006} show
that the chains are far from half-filling and are therefore unlikely
to be strongly correlated, in contrast to the CuO$_2$ planes.  
The metallic nature of the chains is inferred, primarily, from
transport and a.c.\ conductivity measurements.\cite{Basov2005} 

The pairing mechanism for chain superconductivity is not well
established.  Penetration depth anisotropy measurements have also
demonstrated that the chains become superconducting at the same
transition temperature as the CuO$_2$ planes.\cite{Bonn1994} Given the
significant differences in band structure between the two, the most
natural explanation for the single transition temperature is that
chain superconductivity arises from the proximity effect, mediated by
single-electron hopping between the chains and planes.  A generic
feature of $\mathrm{YBa_2Cu_3O_{7-\delta}}$ proximity models is the
presence of a small pairing energy scale associated with chain
superconductivity, which is in addition to the large energy scale
associated with pairing in the CuO$_2$ planes.  The small energy scale
manifests itself, for example, as an inflection point in the
temperature dependence of the superfluid
density.\cite{Atkinson1995,ODonovan1997} The absence of this feature
in microwave experiments on $\mathrm{YBa_2Cu_3O_{7-\delta}}$
originally appeared to indicate a failure of the proximity
model,\cite{Atkinson1995} but has since been shown to be consistent
with the fact that a fraction $\delta$ of oxygen sites are vacant in
the CuO chains.\cite{Atkinson1999} More recently, $\mu$SR experiments
on $\mathrm{YBa_2Cu_3O_{7-\delta}}$ have found an inflection
point,\cite{Sonier2007,Khasanov2007b} but the clearest evidence for
proximity coupling of the chains comes from recent $\mu$SR experiments
on $\mathrm{YBa_2Cu_4O_{8}}$ where there is no chain
disorder.\cite{Khasanov2007}

The goal of our work is to demonstrate that the low-field vortex core
contraction in $\mathrm{YBa_2Cu_3O_{7-\delta}}$ is consistent with the
$\mathrm{YBa_2Cu_3O_{7-\delta}}$ proximity model for chain
superconductivity and is not due to unconventional mechanisms related
to strong correlations (for example, doping-dependent vortex core
expansion in underdoped $\mathrm{La_{2-x}Sr_xCuO_4}$ has been
attributed to coexisting antiferromagnetism\cite{Kadono2004}).  This
work is part of a broader effort to understand how the CuO chains
influence various electronic properties, motivated first by the
possibility of novel physics associated with having a metallic
one-dimensional system coupled to a strongly-correlated
superconductor, and second by a desire to separate the effects of
chains from physics related to strong correlations.

The idea that the different gap energies in multiband superconductors
should introduce distinct magnetic field scales has been explored in
theoretical models for
MgB$_2$,\cite{Nakai2002,Koshelev2003,Dahm2003,Ichioka2004,Zhitomirsky2004,Nicol2005}
and $\mathrm{YBa_2Cu_3O_{7-\delta}}$.\cite{Whelan2000} In particular,
it appears that in these materials the magnetic field dependence of
the density of states (DOS) and related properties such as the
specific heat can be understood if one accounts for the presence of
both a large and a small superconducting gap.  Furthermore, the idea
that the core size in different bands should depend on the gap in each
band has been explored in Refs.~[\onlinecite{Nakai2002,Koshelev2003}],
although an explicit calculation demonstrating vortex core contraction
at low fields has not, to our knowledge, been made.

A brief description of the $\mathrm{YBa_2Cu_3O_{7-\delta}}$ proximity model, followed
by derivations of the appropriate Bogoliubov-deGennes equations, are
presented in Sec.~\ref{theory}.  Results of the calculations are given
in Sec.~\ref{results}, including the main result that the observed
vortex core shrinkage is indeed consistent with the proximity model
for chain superconductivity.  The results are discussed in a broader
context in Sec.~\ref{discussion}, and a brief concluding statement is
made in Sec.~\ref{conclusions}.

\section{Theory}
\label{theory}
In this section we derive the Bogoliubov-deGennes (BdG) equations
appropriate for the proximity model of superconductivity in
$\mathrm{YBa_2Cu_3O_{7-\delta}}$.  Our derivation is similar to ones
described, for example, in Refs.~[\onlinecite{Ghosal2002,Knapp2005}]
but with the additional complications of multi-band superconductivity.
Proximity models for $\mathrm{YBa_2Cu_3O_{7-\delta}}$ have been
discussed in detail elsewhere and we refer the reader to
Refs.~[\onlinecite{Atkinson1995,Atkinson1999}] for more extensive
discussions.

\begin{figure}[tb]
\begin{center}
\includegraphics[width=\columnwidth]{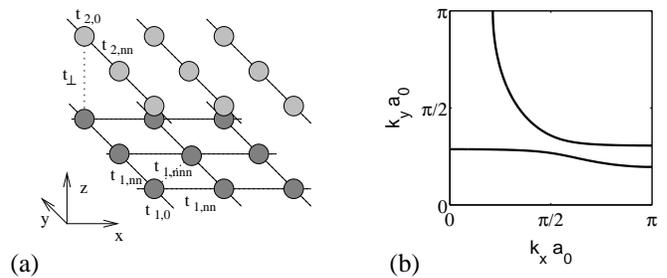}
\caption{ Structure of the bilayer model.  (a) The model consists of a
single plane-chain bilayer with on-site energies $t_{j,0}$ ($j=1,2$)
and single-electron hopping matrix elements along nearest-neighbor
($t_{j,nn}$) and next-nearest-neighbor ($t_{j,nnn}$) bonds as
indicated.  (b) The model has two bands, with the Fermi surfaces as
shown.  The model parameters are
$\{t_{1,0},t_{1,\mathrm{nn}},t_{1,\mathrm{nnn}}\} = \{1,-1,0.45\}$ for
the plane layer, $\{t_{2,0},t_{2,\mathrm{nn}}\} = \{2.4,-2\}$ for the
chain layer, and the interlayer hopping amplitude is $t_\perp = 0.3$.
The pair interaction in the plane layer is $V=1.6$, which produces a
zero-field pair amplitude $\Delta = 0.39$ for the single-layer model
and $\Delta = 0.33$ for the bilayer model.  All energies are in units
of $|t_{1,\mathrm{nn}}|$.}
\label{fig:fs}
\end{center}
\end{figure}

The geometry of the plane-chain model is illustrated in
Fig.~\ref{fig:fs}.  The model consists of a single bilayer, comprising
a two-dimensional layer (aligned with the $x$-$y$ plane) and a layer
of one-dimensional chains (aligned with the $y$-axis).  The 2D plane
represents a CuO$_2$ layer and is coupled via single-electron hopping
to the chain layer.  This is the simplest model that contains the
essential physics of multi-band superconductivity in
$\mathrm{YBa_2Cu_3O_{7-\delta}}$.  There is an intrinsic pairing interaction $V$ in
the plane, but the chains are intrinsically normal, which means
that the superconducting order parameter is nonzero in the plane layer
only.  The chains are, nonetheless,
superconducting and exhibit a gap in their DOS.  One important feature
of this model that distinguishes $\mathrm{YBa_2Cu_3O_{7-\delta}}$ from other
multiband superconductors is that the hybridization of the plane and
chain layers is strongly $k$-dependent and, consequently, the induced
gap in the chain layer does not have a simple $d$-wave symmetry, even
though the order parameter does.\cite{Atkinson1999,s-wave} A consequence of
this is that the chain-projected DOS exhibits two pairs of coherence
peaks,\cite{Atkinson1995,Atkinson1999,Whelan2000,Ngai2007} meaning
that there is more than one superconducting energy scale in the
chains.  We refer to these as the ``small'' and ``large'' energy
scales.

We consider only magnetic fields aligned with the crystalline
$c$-axis (which we align with the $z$-axis), perpendicular to the
CuO$_2$ plane, such that screening currents circulate within the plane
and chain layers.  It is for this configuration that the vortex core
contraction is seen in $\mu$SR experiments.\cite{Sonier1999b}  
The total Hamiltonian for the model can be broken into pieces 
\begin{equation}
\hat H = \hat H_1 + \hat H_2 + \hat H_\perp.
\end{equation}
where $\hat H_1$ is the Hamiltonian for the isolated plane, 
$\hat H_2$ the Hamiltonian for the isolated chains, and 
$\hat H_\perp$ the single-electron hopping term that couples the two layers.
For comparison, we also consider a single-layer model described by
$\hat H_1$ alone.

The BdG Hamiltonian for the isolated plane is
\begin{eqnarray}
\hat H_1 &=& \sum_{ij\sigma} \tilde t_{1ij}
c^\dagger_{1\sigma}(\rr_i) c_{1\sigma}(\rr_j) 
+ \sum_{ij} [
\Delta_{ij} c^\dagger_{1\uparrow}(\rr_i)
c^\dagger_{1\downarrow}(\rr_j) 
\nonumber \\ && 
\qquad +\Delta_{ij}^\ast
c_{1\downarrow}(\rr_j) c_{1\uparrow}(\rr_i) ],
\end{eqnarray}
where $c_{1\sigma}(\rr_i)$ is the annihilation operator for an
electron in the plane on site $i$ with spin $\sigma$, and position
$\rr_i=(x_i,y_i)$, $\tilde t_{1ij}$ are hopping matrix elements, and
$\Delta_{ij}$ are superconducting pair energies.  The subscripts ``1''
and ``2'' refer to the plane and chain layers
respectively.  The hopping matrix element $\tilde t_{1ij}$ between
sites $i$ and $j$ includes the effects of the magnetic field via the
Peierls substitution:
\begin{eqnarray}
\tilde t_{1ij} &=& 
t_{1ij}e^{-i(e/\hbar c)\int_{\rr_j}^{\rr_i} d\rr\cdot
{\bf A}(\rr)} \nonumber \\
&=& t_{1ij} e^{\left[ i\alpha \frac{y_i+y_j}{2} (x_i-x_j) \right ]},
\label{Eq:tij}
\end{eqnarray}
where $t_{1ij}$ are the zero-field matrix elements.  Here ${\bf
A}(\rr_i) = -B_0y_i{\bf \hat x}$ is the static magnetic vector
potential, where $B_0$ is the uniform applied magnetic field and
\begin{equation}
\alpha = eB_0/\hbar c.
\end{equation}  
In principle, the inhomogeneous 
magnetic field ${\bf B}(\rr) = B_0 {\bf \hat z} + \delta {\bf B}(\rr)$ 
should be calculated self-consistently and the hopping matrix elements in
Eq.~(\ref{Eq:tij}) modified accordingly.  However, our calculations are
performed for large fields where, as we show below, $\delta {\bf B}(\rr)$
is small and can be neglected.

We take a second-nearest neigbor model with zero-field matrix elements
$t_{1ii} = t_{1,0}$, $t_{1\langle i,j\rangle}=t_{1,\mathrm{nn}}$, and
$t_{1\langle \langle i,j \rangle \rangle} = t_{1,\mathrm{nnn}}$, where
$\langle i,j\rangle$ and $\langle\langle i,j\rangle \rangle$ refer to
nearest and next-nearest neighbors respectively (cf.\
Fig.~\ref{fig:fs}).  In the zero-field limit, the dispersion for the
plane layer is $\epsilon_1(\kk) = t_{1,0} + 2t_{1,\mathrm{nn}}( \cos
k_x + \cos k_y) + 4t_{1,\mathrm{nnn}}\cos k_x \cos k_y$.

The local superconducting order parameter $\Delta_{ij}$ is determined
self-consistently under the assumption that the pair interaction $V$ is
attractive for nearest neighbor electrons but vanishes otherwise.
Then, 
\begin{equation}
 \Delta_{ ij } = -\frac {V}{2} \langle
c_{1\downarrow}(\rr_j) c_{1\uparrow}(\rr_i) + c_{1\downarrow}(\rr_i)
c_{1\uparrow}(\rr_j) \rangle \delta_{\langle i,j \rangle}.
\end{equation}
 The d-wave component, defined by
\begin{equation}
\Delta(\rr_i) = \sum_j (-1)^{y_i-y_j}\Delta_{ij},
\label{Eq:Delta}
\end{equation} 
is the dominant component of the order parameter.

The isolated chain layer is described by a Hamiltonian 
\begin{equation}
H_2 = \sum_{ij\sigma} t_{2ij} c^\dagger_{2\sigma}(\rr_i)
c_{2\sigma}(\rr_j)
\end{equation}
where $t_{2ii}=t_{2,0}$ and $t_{2ij} = t_{2,\mathrm{nn}}$ for $i$ and
$j$ nearest-neighbor sites belonging to the same chain.  Note that,
because of our choice of gauge, the hopping matrix elements are
unchanged by the magnetic field.  The zero-field dispersion for the
chains is $\epsilon_2(\kk) = t_{2,0} + 2t_{2,\mathrm{nn}}\cos k_y$.
The layers are coupled by interlayer hopping:
\begin{equation}
H_\perp = t_\perp \sum_{i\sigma}
[c_{1\sigma}^\dagger(\rr_i)c_{2\sigma}(\rr_i) + c_{2\sigma}^\dagger(\rr_i)
c_{1\sigma}(\rr_i)],
\end{equation}
 which mixes the chain and plane wavefunctions.  Rather than attempt a
quantitative description of $\mathrm{YBa_2Cu_3O_{7-\delta}}$, we choose band parameters (cf.\
Fig.~\ref{fig:fs}) which are optimal for numerical calculations, but
which preserve the general features of the $\mathrm{YBa_2Cu_3O_{7-\delta}}$ Fermi surface.

While the Hamiltonian is not periodic, there is nonetheless a
quasi-periodicity which allows us to define an $L_x\times L_y$
magnetic supercell containing $N = L_xL_y/a_0^2$ atomic lattice sites
($a_0$ is the lattice constant) and enclosing an even number of
 flux quanta, where the superconducting flux quantum is $\Phi_0 \equiv hc/2e$.
We take two vortices per supercell so that
\begin{equation}
B_0 = \frac{2\Phi_0}{L_xL_y}.
\end{equation}
Assuming there are $N_k = N_{kx} N_{ky}$ supercells in the system, we
can define Bloch states via the transformation
\begin{equation}
c_{n i\KK \sigma} = \sum_{I=1}^{N_k} 
c_{n\sigma}({\rr}_i+\RR_I)
\frac{e^{-i(\KK\cdot\RR_I + \alpha x_i Y_I)}}{\sqrt{N_k}} 
\end{equation}
where $\RR_I=(X_I,Y_I)$ are the supercell lattice vectors labelled by
$I$, and where $\rr_i=(x_i,y_i)$ now, and hereafter, labels
 sites with site index $i\in
[1,N]$ within the magnetic supercell.  The supercell wavevector is
$\KK = {2\pi( n_x,n_y )}/{L}$,
where $L=N_{kx}L_x=N_{ky}L_y$ is the linear dimension of the system
and $n_x$ and $n_y$ are integers such that $n_x \in [1,N_{kx}]$,
$n_y \in [1,N_{ky}]$.
The Hamiltonian is block-diagonal in this new basis, and has the form
\begin{equation}
\hat H = \sum_\KK \sum_{ij} \hat \Psi^\dagger_i(\KK) H_{ij}(\KK)
\hat \Psi_j(\KK)
\end{equation}
with
\begin{equation}
H_{ij}(\KK) = 
\left[ \begin{array}{cccc} 
    \tilde t_{1ij}(\KK) & \Delta_{ij}(\KK) & t_\perp & 0 \\
    \Delta^\dagger_{ij}(\KK) & -\tilde t_{1ij}(-\KK)^\ast & 0 & -t_\perp \\
    t_\perp & 0 &     \tilde t_{2ij}(\KK) & 0\\
    0 & -t_\perp & 0 & -\tilde t_{2ij}(-\KK)^\ast \\
  \end{array} 
  \right ],
\end{equation}
where
 $ \hat \Psi_i^\dagger(\KK) = [c^\dagger_{1 i\KK\uparrow},
c_{1 i\,-\KK\downarrow}, c^\dagger_{2 i\KK\uparrow},
c_{2 i\,-\KK\downarrow}]$ and
\begin{equation}
\tilde t_{nij}(\KK) =  t_{nij} e^{-i\KK\cdot \RR}
e^{i\alpha \left \{ \frac{y_i+y_j}{2}(x_i-x_j + X) - \frac{x_i+x_j}{2}Y 
+ \frac{XY}{2}\right \} }.
\end{equation}
The hopping matrix elements $t_{nij}(\KK)$ have periodic boundary
conditions at the edges of the supercell: an electron at $\rr_j$ which
leaves the supercell via one of its edges is periodically mapped back
onto site $\rr_i$ belonging to the supercell via the appropriate
supercell lattice vector $\RR = (X,Y)$.

In terms of Bloch states, the superconducting gap amplitude is
\begin{eqnarray}
\Delta_{ij}(\KK) 
&=& \frac{1}{N_k} \sum_{\KK^\prime} V_{ij}(\KK-\KK^\prime) \nonumber \\
&&\times
\langle c_{1j\,-\KK^\prime \downarrow}
c_{1i\KK^\prime \uparrow} +
c_{1i\KK^\prime \downarrow} c_{1j\,-\KK^\prime \uparrow}
\rangle,
\end{eqnarray}
with 
$V_{ij}(\qq) = -\frac{1}{2}V \delta_{\langle i,j\rangle} e^{i\qq\cdot\RR}$.
In many experiments, in particular $\mu$SR, it is not the
order parameter but the magnetic field profile which is measured near
a vortex core.  This is directly related to the  current densities in
the chain and plane layers.  The 2D current density 
at a site $i$ in layer $n$ is defined by averaging the current densities
flowing into and away from the site, 
\begin{eqnarray}
{\bf J}^{2D}_n(\rr_i)
&=& \frac{-e}{2 \hbar a_0} \mbox{Im} \sum_{\sigma,\KK,j}
\delta \rr \,\tilde t_{ij}(\KK) \langle c^\dagger_{n,i\KK\sigma}
c_{n,j\KK\sigma} \rangle,
\label{eq:current}
\end{eqnarray}
where the prefactor $\frac 12$ comes from the average and $\delta {\bf
r} = \rr_i - \rr_j + \RR$.

We remark that the calculations described above are gauge invariant
provided $N_{kx} = L_y/a_0$ and $N_{ky} = L_x/a_0$.  In practice, it
is not feasible to sum over such a large number of k-points
 at low fields where $L_x$ and $L_y$ are large, and by
necessity we use a reduced set at the lowest field strengths.  A
consequence of this approximation is that the current in the normal
state does not vanish identically.  This is particularly problematic
for the chain layer.  For system sizes up to $L_{x,y}=14a_0$ no
approximation is made, while for systems up to $L_{x,y}=40a_0$,
$N_{kx,ky} = L_{y,x}/2a_0$.  For the largest system sizes,
$L_{x,y}=50a_0$ and $L_{x,y}=60a_0$, we have taken $N_{kx,ky} = 5$.  We
have checked that the spurious normal-state current in the largest
systems is at least an order of magnitude smaller than the currents
reported here in the superconducting state.

\begin{figure}[tb]
\begin{center}
\includegraphics[width=\columnwidth]{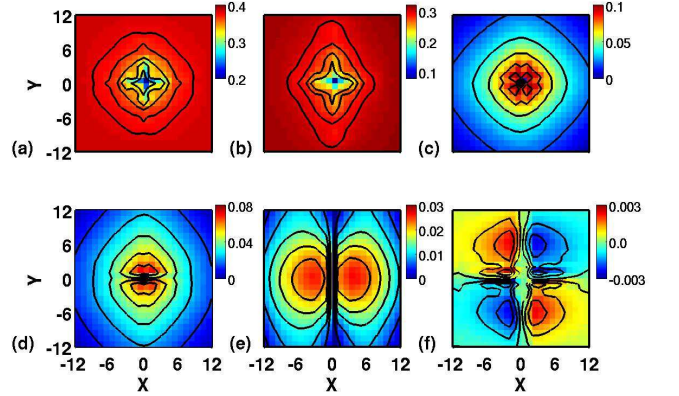}
\end{center}
\caption{(color online) Self-consistent solutions for the vortex structure
(diamond lattice) with applied field $B_0 = 2\Phi_0/2500a_0^2$.  The
d-wave order parameter is shown as a function of position (in units of
$a_0$) for (a) the single-layer and (b) bilayer models.  The current
magnitudes are shown for (c) the single-layer and (d-f) the bilayer models.
For the bilayer the plane (d), chain (e), and interlayer (f) current
amplitudes are shown.}
\label{fig:scsoln}
\end{figure}

\section{Results}
\label{results}
In this section we describe the results of self-consistent
solutions of the BdG equations for the vortex lattice.  Our goal is to
explain the observed magnetic field dependence of the core size at low
fields.  We have performed calculations for diamond, square, and
monoclinic lattice structures and have found the same qualitative
results for the vortex core size in all cases.  We present results
for the diamond lattice, for which $L_x = L_y = L$ and the lattice
vectors for the primitive unit cell are $(L/2,\pm L/2)$.

The self-consistently determined d-wave gap and current distributions
are shown in Fig.~\ref{fig:scsoln} near a single vortex.  Note that
the vortex cores shown in Figs.~\ref{fig:scsoln}(a,b) have a radius of
roughly $2a_0$, whereas the coherence length is $\sim 5a_0$ in
optimally-doped YBCO$_{6.95}$.\cite{Sonier2007} This discrepancy
results from our having taken the order parameter to be twice what is
appropriate for quantitative models of
$\mathrm{YBa_2Cu_3O_{7-\delta}}$.  We have done this so that energy
scale of the the induced gap in the chain layer lies near the middle
of the range of numerically accessible magnetic fields.

Figure~\ref{fig:scsoln} illustrates the various effects of proximity
coupling on the vortex structure.  First, there is an overall
reduction of the order parameter in the plane layer owing to the
presence of the chains. This is a general feature of proximity models
which is independent of the magnetic field: while the plane induces
superconductivity in the chains, the (intrinsically-normal) chains
also degrade superconductivity in the plane.  Similar physics has been
found in multiband models for MgB$_2$ where impurity scattering
between the $\pi$-band and $\sigma$-band mixes the two
bands.\cite{Nicol2005} In our model, this mixing comes from the
interlayer hopping and means that the vortex core in the bilayer model
is slightly larger than for the single layer model.

Second, we note that there is an anisotropic suppression of the order
parameter near the vortex cores which is evident in the bilayer model,
with the cores being extended along the chain direction.  This follows
from the anisotropy of the current in the plane,
Fig.~\ref{fig:scsoln}(d), which itself follows from two features of
the proximity model: (i) the chains and plane carry currents in
parallel, and (ii) the chains only carry currents in the $\hat y$
direction.  The current in the plane is
consequently larger at positions along the $y$-axis, where it flows
entirely in the $\hat x$ direction [and the chain-current therefore vanishes,
Fig.~\ref{fig:scsoln}(e)], than at corresponding positions along the
$x$-axis.

Third, there is an interlayer current, Fig.~\ref{fig:scsoln}(f), that
has a quadropolar structure and introduces a small quadropolar
in-plane component to the magnetic field.  Since the interlayer current density
is an order of magnitude smaller than the intralayer current density,
the in-plane component is small.

We wish to extract a vortex core size from our calculations; however,
the vortex core is not a well-defined object and the core size is not
uniquely defined.  In simple BCS superconductors, the various
definitions give similar results\cite{Sonier2004b} but the situation
is more complicated in $\mathrm{YBa_2Cu_3O_{7-\delta}}$ where there is more than a
single length scale.  The most obvious measure of the vortex core size
is the length scale over which the order parameter approaches its
asymptotic value, usually determined from the gradient of the order
parameter near the vortex core center.\cite{Sonier2004b} This
definition does not work well when the coherence length and the
lattice constant are comparable, as we have here.  Similarly, the
commonly-used definition that the core size is given by the radius at
which the current density is a maximum suffers from poor resolution
due to the discreteness of the atomic lattice.  In the following, we discuss
a procedure for extracting the core size from the vorticity of the current
distribution.

\begin{figure}[tb]
\begin{center}
\includegraphics[width=\columnwidth]{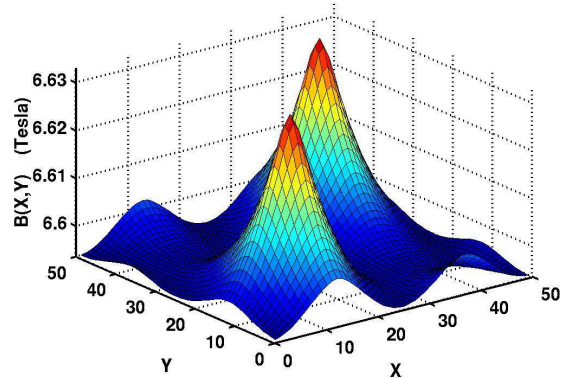}
\end{center}
\caption{(color online) Magnetic field for a single magnetic unit cell
containing two vortices.  Results are shown for the bilayer model with
$B_0 = 2\Phi_0^2/2500a_0^2$.  The calculations assume that $a_0 = 5$ \AA,
$d_z = 10$ \AA, and that the energy scale is $|t_{1,\mathrm{nn}}|=100$
meV.}
\label{fig:Bfield}
\end{figure}

This approach is motivated by the fact that 
$\mu$SR experiments measure the distribution of the magnetic field in
the vortex lattice.  The magnetic field inhomogeneity $\delta {\bf B}
(\rr) = {\bf B}(\rr) - {\bf B_0}$ can be calculated from the current
density profiles via Maxwell's equation $\nabla \times \delta {\bf B}
= (4\pi/c){\bf J}$, where ${\bf J}(\rr)$ is the volume current
density.
The $z$-component of $\delta{\bf B}(\rr)$ satisfies
\begin{equation}
\nabla^2 \delta B_z(\rr) = -\frac{4\pi}{c} (\nabla\times {\bf J})\cdot
{\bf \hat z},
\label{eq:B}
\end{equation}
which can be solved using a Jacobian relaxation scheme.  For
illustrative purposes, we make the approximation that the small interlayer
currents can be neglected and that the current is uniformly
distributed along the c-axis, i.e. that ${\bf J}(\rr) = [{\bf
J}_1^{2D}(x,y) + {\bf J}_2^{2D}(x,y)]/d_z$ is independent of $z$,
where $d_z$ is the c-axis lattice constant for the atomic unit cell.
In order to extract quantitative
values, we adopt approximate model parameters for
$\mathrm{YBa_2Cu_3O_{7-\delta}}$ (see caption of Fig.~\ref{fig:Bfield}).   
The plot of $B_z(\rr)$ shown Fig.~\ref{fig:Bfield} is for a magnetic
field near the lower limit of what is computationally accessible.
We note that the field varies by $\approx 0.5\%$
over the magnetic unit cell, which justifies the approximation made in
Eq.~(\ref{Eq:tij}) that the field is uniform.

\begin{figure}[tb]
\includegraphics[width=\columnwidth]{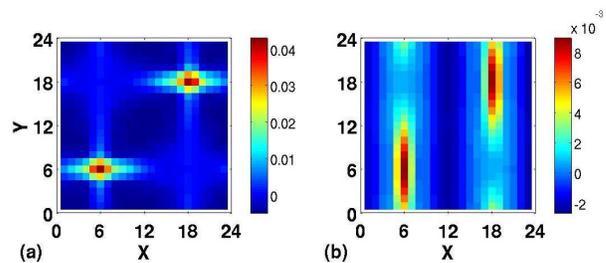}
\caption{(color online) Vorticity $\omega(\rr)$ of superfluid currents
in (a) the plane layer and (b) the chain layer for the bilayer model
at $B_0 = 2\Phi_0/576a_0^2$.}
\label{fig:curlj}
\end{figure}

One can extract a core size from $\delta {\bf B}(\rr)$, for example by
fitting to a Ginzburg-Landau form for the vortex
lattice;\cite{Yaouanc1997} however, we note from Eq.~(\ref{eq:B}) that
the vorticity 
\begin{equation}
\mathbf{\omega}(\rr) \equiv \nabla\times {\bf J}(\rr),
\end{equation}
gives the vortex core size directly.  For the simple example of an
isolated vortex in an isotropic medium, ${\bf J} \sim {\bf \hat\theta}
\tanh^2(r/\xi)/r$, and $\mathbf{\omega}(r)$ decays exponentially for
$r > \xi$.  One can then extract a characteristic vortex core size
from the second moment of the position along the $x$-axis since
\[
\langle x^2 \rangle =  \frac{\int d^2r\,
\omega(\rr) x^2 }
{\int d^2r\, \omega(\rr)} = 0.693147 \xi^2.
\]

In a vortex lattice, the second moment is not well defined 
since ${\bf \omega}(\rr)$ satisfies
\begin{equation}
\int  d^2 r\, {\bf \omega}(\rr) = 0,
\end{equation}
(where the integral is over the vortex unit cell) and is therefore not
positive definite.  However, if the circulation around each vortex is
positive (negative), then we can still extract a characteristic core
size based on the region over which ${\bf \omega}(\rr)$ is positive
(negative).  We define the extent of the vortex $\rho_{\bf \hat n}$
along a direction ${\bf \hat n}$ as
\begin{equation}
\rho_{\bf \hat n} \equiv \left [ \frac{\int_{\omega>0} d^2r\,
\omega(\rr)( \rr\cdot{\bf \hat n} )^2}
{\int_{\omega>0} d^2r\, \omega(\rr)} \right ]^{1/2}
\end{equation}
where the integral is taken over a single vortex.  This definition is
not unique, but it serves to illustrate the physics of the vortex core
contraction at low fields. 
The main advantage of this approach is that it is relatively
insensitive to the discreteness of the lattice.

Examples of the vorticity are shown in Fig.~\ref{fig:curlj} for the
plane and chain layers.  One sees that the vortex core in the chain layer is
highly anisotropic, and is extended along the $y$-direction.
\begin{figure}
\begin{center}
\includegraphics[width=\columnwidth]{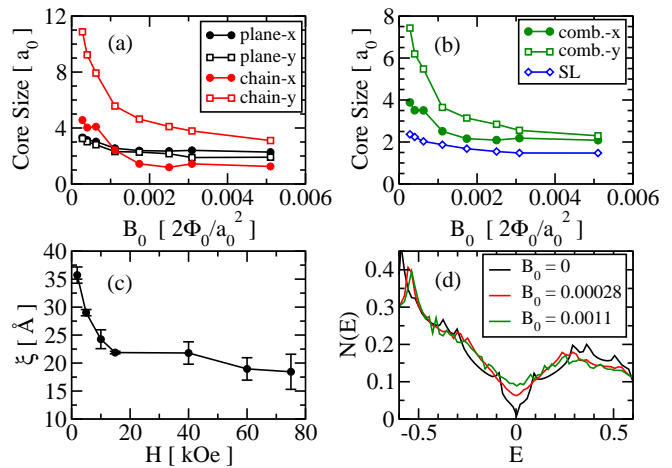}
\end{center}
\caption{(color online) Core size for the bilayer model along the $x$
and $y$ directions for (a) the plane and chain layers and (b) for the
combined current distribution.  The core size for the single-layer
(SL) model is also shown.  (c) Experimental data from
Ref.~[\onlinecite{Sonier2007}] is shown for comparison. (d) The
chain-projected density of states for the bilayer model is plotted for
different fields ($B_0$ is in units of $2\Phi_0/a_0^2$). }
\label{fig:coresize}
\end{figure}
In Fig.~\ref{fig:coresize}(a), we plot the field-dependence of the
core sizes $\rho_{\bf \hat x}$ and $\rho_{\bf \hat y}$ for the plane
and chain layers based on their separate current distributions ${\bf
J}_1^{2D}(\rr)$ and ${\bf J}_2^{2D}(\rr)$ respectively.  The main
point of this figure is that while the core size in the plane layer
depends only weakly on the field, $\rho_{\bf \hat y}$ (and to a lesser
extent $\rho_{\bf \hat x}$) in the chain layer varies rapidly with
$B_0$ for $B_0 < B^\ast$.  Empirically, 
\begin{equation}
B^\ast \approx 0.001 (2\Phi_0/a_0^2)
\label{bstar}
\end{equation}
for the model parameters used in this work.  In
Fig.~\ref{fig:coresize}(b), we show similar calculations for
$\rho_{\bf \hat x}$ and $\rho_{\bf \hat y}$ based on the combined
current ${\bf J}^{2D} = {\bf J}_1^{2D} + {\bf J}_2^{2D}$.  Again,
there is a rapid core contraction with increasing $B_0$, primarily
along the $y$-direction, for $B_0 < B^\ast$.
Experimental measurements of the core size\cite{Sonier2007}, plotted
in Fig.~\ref{fig:coresize}(c), show a similar variation at low field.  In
comparison, there is a relatively weak low-field core-size variation for
the single-layer model [Fig.~\ref{fig:coresize}(b)], a factor of about
1.6 over the range of $B$ shown, in quantitative agreement with
earlier calculations.\cite{Ichioka1999a} Figure~\ref{fig:coresize}(b)
is the main result of this work.

We note that the results in Fig.~\ref{fig:coresize} are in qualitative
agreement with a simplified quasiclassical ``doppler-shift''
calculation that has been reported previously\cite{Sonier2007}.  The
current results confirm the validity of the previous approximate calculations.

The density of states (DOS) for the chain layer
[Fig.~\ref{fig:coresize}(d)] shows that there are two distinct 
superconducting energy scales in the chain spectrum, 
a large gap $E_L \approx 0.35$ and a small gap $E_S \approx 0.1$.  The
two-gap spectrum originates from the one-dimensional structure of the
chains, and is discussed at length in
Refs.~[\onlinecite{Atkinson1995,Atkinson1999}].  In the range of
fields explored (which are much lower than the upper critical field),
the chain DOS in the interval $|E|<E_S$ is a strong function of field
for $B_0 < B^\ast$ but saturates for $B_0 > B^\ast$.  This illustrates
the close connection between $E_S$ and $B^\ast$.  We expect that
$B^\ast$ is the field at which the vortex cores in the chain layer
begin to overlap.  We estimate a BCS length scale
$\xi_{\textit{chain}} =\hbar v_{F,\textit{chain}}/\pi E_S$ for the
small gap, where $v_{F,\textit{chain}}$ is the $y$-component of the
Fermi velocity in the chain.  For our model parameters this gives
$\xi_{\textit{chain}} \sim 10a_0$, which is close to the low-field
value of $\rho_{\bf \hat y}$ for the chain layer shown in
Fig.~\ref{fig:coresize}(a).  For the diamond lattice, the vortex
spacing along the chain direction is $\sqrt{2}\ell_m$, where $\ell_m
=\sqrt{\Phi_0/B_0}$ is the magnetic length.  Then, the vortex cores
will overlap when $\sqrt{2}\ell_m \approx 2\xi_{\textit{chain}}$,
which gives an estimate for $B^\ast$ of
\begin{equation}
B^\ast \approx \frac{\Phi_0}{2\xi_{\textit{chain}}^2}.
\label{bstar2}
\end{equation}
For the current model, this gives $B^\ast \approx 0.0025 (2\Phi_0/a_0^2)$,
in good agreement with Eq.~(\ref{bstar}).

\section{Discussion}
\label{discussion}
In this section, we discuss our results in the context of related published
work.

A number of tunneling experiments on
$\mathrm{YBa_2Cu_3O_{7-\delta}}$\cite{Sun1994,Ngai2007} have found a
spectrum with multiple superconducting energy scales.  While the
origin of the different scales has not been firmly established, there
is evidence that they arise from a single pairing interaction,
consistent with the proximity model.\cite{Ngai2007} The smallest of
the measured gaps is $\sim 5$ meV in YBCO$_{6.95}$ and it is believed
to arise from chain superconductivity.  If we then take $\hbar
v_{F,\textit{chain}} = 4.12$ eV\AA{} from first principles band
structure calculations\cite{Andersen1995}, then we get
$\xi_{\textit{chain}} = 262$ \AA.  This, using Eq.(\ref{bstar2}),
gives a crossover field of $B^\ast \sim 1.5$~T, which is in remarkably
close agreement with experimental measurements reproduced in
Fig.~\ref{fig:coresize}(c).  This has two implications.  First, it
appears to indicate consistency between two distinct experiments, one
of which (tunneling) is surface sensitive.  Second, it strengthens the
case that the proximity model is  appropriate for
$\mathrm{YBa_2Cu_3O_{7-\delta}}$.

One of the key features of the $\mathrm{YBa_2Cu_3O_{7-\delta}}$
proximity model is that the pairing interaction resides within the
plane layer and that pairing in the chains results from
single-electron hopping between the physical layers.  In this model,
the induced gap in the chain layer is proportional to the gap in the
planes.  A similar model has been introduced for
MgB$_2$:\cite{Dahm2003,Ichioka2004} a domininant intraband pairing
interaction in the $\sigma$-band is assumed, and a subdominant pairing
interaction in the $\pi$-band arises through interband pair-tunneling.
There is a qualitative similarity in the field dependence of the DOS
between the two models: in both cases, the low-energy DOS fills in
rapidly as the field increases, but the energy of the gap edge is
nearly field-independent [Fig.~\ref{fig:coresize}(d)].  This
should be contrasted with single-band superconductors where 
the gap is field-dependent.\cite{Millstein1967}
In $\mathrm{YBa_2Cu_3O_{7-\delta}}$, field-induced pair
breaking occurs primarily in the chain layer, but the pairing
interaction resides in the plane layer.

There are also important physical distinctions between the MgB$_2$ and
$\mathrm{YBa_2Cu_3O_{7-\delta}}$ models.  In MgB$_2$ models, pairing
is generally assumed to occur in the short-ranged s-wave channel.
Thus, Cooper pairs belong either entirely to the $\sigma$-band or
$\pi$-band.  In contrast, a significant contribution to chain
superconductivity in the $\mathrm{YBa_2Cu_3O_{7-\delta}}$ model comes
from pairing correlations between electrons in the plane and chain
layers.\cite{Atkinson1999} Furthermore, the predominantly $d$-wave
symmetry of the order parameter in the plane layer {\em cannot} imply
a fourfold-symmetric chain gap because the underlying chains are
one-dimensional.  Thus, unlike in MgB$_2$, the $\kk$-resolved
excitation spectrum of the chains is quite complicated and there is no
single gap energy that one can attach to chain superconductivity.

One consequence of this is that, since $B^\ast$ is associated with the
smallest of the chain gaps, a significant superfluid density remains
in the chain layer when $B_0 > B^\ast$.  Even at the largest field
studied, the chain DOS at the Fermi energy is about half its value in
the normal state [cf.\ Fig.~\ref{fig:coresize}(d)], meaning that
Cooper pairs formed with binding energies corresponding to the large
gap are not broken by the magnetic field.

One interesting question, which is beyond the scope of this work, is
how the structure of the vortex lattice itself is affected by the
presence of the chains.  It has been found that, at fields less than
4~T, the vortex lattice in $\mathrm{YBa_2Cu_3O_{7-\delta}}$ has a
distorted hexagonal symmetry\cite{Brown2004} with the distortions
apparently originating from the CuO chains.\cite{Johnson1999} At
higher fields there is a crossover to a square lattice, which is
expected from the $d$-wave symmetry of the order parameter in the
CuO$_2$ planes.  These experiments suggest that chain
superconductivity is degraded for $B> 4$~T. While this field is
approximately 2.5 times the value of $B^\ast$ we extracted by eye from
the data in Fig.~\ref{fig:coresize}(c), we suggest that the crossover in the
vortex lattice structure is generally consistent with both the 
$\mu$SR measurements and the proximity model.  Quantitative calculations
are needed to establish rigorous consistency.

Finally, we note that while the calculations in this work assume that
the chains are infinitely long, $\mathrm{YBa_2Cu_3O_{7-\delta}}$ has a fraction
$\delta$ of chain-layer oxygen sites which are vacant.  O-vacancies
effectively divide the chains into fragments of varying lengths
$\ell$, and it is worth considering how this affects the results
presented here.  Based on our earlier assertion that $B^\ast$ is the
field at which vortex cores in the chain layer overlap, we suggest
that the low-field vortex core-size variation should be easily observable
provided that the mean value $\overline \ell$ of $\ell$ is larger than
$2\xi_\textit{chain}\sim 500$\AA.  The experiments of
Ref.~[\onlinecite{Sonier2007}] span a range of fillings between
YBCO$_{6.57}$ and YBCO$_{6.95}$ so it is possible that there are large
sample-to-sample variations in $\overline \ell$.  For
randomly-distributed O-vacancies, $\ell=1/\delta$; however it is well
known that O-vacancies cluster after annealing and that the chain
fragments are typically much longer. In YBCO$_{6.5}$, for example, the
chains alternate between being completely filled and completely empty.
In practice, in any sample there will be a distribution of $\ell$,
with some fraction $n_\textit{chain}$ of these satisfying $\ell >
2\xi_\textit{chain}$.  Since the magnitude of the low-field
core contraction depends on the magnitude of the current circulating in
the chain layer, we expect the low-field vortex core size to depend on 
 $n_\textit{chain}$.  On the other hand, the crossover
field $B^\ast$ depends primarily on the magnitude of the
plane-chain coupling and should not be directly dependent on
$n_\textit{chain}$.\cite{complications}  Although data sets for $\delta > 0.05$
in Ref.~[\onlinecite{Sonier2007}] sample a limited set of magnetic
field strengths, they appear consistent with this picture.

\section{Conclusions}
\label{conclusions}
We have studied the vortex core structure within a simple proximity
model for $\mathrm{YBa_2Cu_3O_{7-\delta}}$.  We find that the current distribution
around the vortex core is different in the chain and
plane layers, and that the core is elongated along the chain
direction.  There is a crossover in the magnetic field dependence of
the core size at a field $B^\ast$.  The core size varies rapidly with
magnetic field $B_0$ for $B_0 < B^\ast$, and we have shown that
$B^\ast$ is related to the energy scale of the small superconducting
gap in the chain layer.  Our calculations provide a natural explanation
for the vortex core contraction measured in various $\mu$SR experiments, and
support the validity of the proximity model for $\mathrm{YBa_2Cu_3O_{7-\delta}}$.

\section*{Acknowledgments} We acknowledge support by NSERC of Canada, the
Canada Foundation for Innovation, the Ontario Innovation Trust, and
the Canadian Institute for Advanced Research.


\begin{thebibliography}{55}
\bibitem[{com()}]{complications}
\bibinfo{note}{The crossover field $B^\ast$ will depend on the distribution of
  $\ell$ indirectly because the latter affects the doping in the CuO$_2$
  planes.}
\end{thebibliography}

\end{document}